\documentclass
[aps,prl,twocolumn,floatfix,english,showpacs,10pt,superscriptaddress]{revtex4-2}%
\usepackage{graphicx}
\usepackage{booktabs}
\usepackage{amsmath}
\usepackage{physics}
\usepackage{amssymb}
\usepackage{colordvi}
\usepackage{verbatim}
\usepackage{xcolor}
\usepackage{mathrsfs}
\usepackage{epsfig}
\usepackage{lipsum}
\usepackage{amsfonts}
\usepackage{makecell}
\usepackage{esint}
\usepackage{bm}
\usepackage[most]{tcolorbox}

\usepackage[unicode=true, breaklinks=false, pdfborder={0 0 1}, backref=false,
colorlinks=true, linkcolor=blue, urlcolor=blue, citecolor=blue]{hyperref}%
\providecommand{\U}[1]{\protect\rule{.1in}{.1in}}
\providecommand{\U}[1]{\protect\rule{.1in}{.1in}}
\setcitestyle{numbers,square}

\begin{document}

\title{Vector Magnonics: Electrical Injection and Control of Spin Flow in Altermagnets}

\author{Yanmeng Lei}
\affiliation{School of Physics, Huazhong University of Science and Technology, Wuhan 430074, China}

\author{Rui-Chun Xiao}
\affiliation{Institute of Physical Science and Information Technology, Anhui University, Hefei 230601, China}
\affiliation{Anhui Provincial Key Laboratory of Magnetic Functional Materials and Devices, Anhui University, Hefei 230601, China}

\author{Weiwei Lin}
\affiliation{School of Physics, Southeast University, Nanjing 211189, China}

\author{Tao Yu}
\email{taoyuphy@hust.edu.cn}
\affiliation{School of Physics, Huazhong University of Science and Technology, Wuhan 430074, China}

\date{\today }

\begin{abstract}
Altermagnets host chirally split magnons that promise unique functionalities for information processing. However, their distinctive transport signatures, crucial for experimental identification and manipulation, remain elusive. Here, we predict that a spin accumulation electrically injects a ``vector" or multidirectional magnon spin current into an altermagnet, comprising both longitudinal and sizable transverse components. Notably, this transverse current exhibits a sign reversal away from the source and can be switched on or off by reorienting the N\'eel vector. While such a transverse current is found to be not forbidden even in conventional antiferromagnets, we demonstrate through quantum-kinetic calculations that in altermagnets, the transverse response is enhanced by two orders of magnitude due to broken parity-time symmetry. This giant enhancement provides a decisive transport fingerprint for detecting magnon spin splitting and N\'eel-vector orientation, offering a clear criterion to experimentally distinguish altermagnets from conventional antiferromagnets.
\end{abstract}

\maketitle

Altermagnets (ATMs) constitute a distinct magnetic phase that combines the high-frequency dynamics of antiferromagnets (AFMs) with the pronounced spin-dependent responses of ferromagnets~\cite{ATM1,ATM2,ATM3,ATM4,ATM5}. A key feature is a unique interplay of crystal and spin-rotation symmetries that breaks global Kramers degeneracy~\cite{ATM2,Symmetry}, producing the observed non-relativistic spin-splitting in electronic bands via X-ray magnetic circular dichroism~\cite{X-ray1,X-ray2,X-ray3} and angle-resolved photoemission spectroscopy~\cite{spin_splitting_1,spin_splitting_2,spin_splitting_3,spin_splitting_4,spin_splitting_5,spin_splitting_6,spin_splitting_7}.

This spin splitting enables novel spintronic effects, including a non-relativistic spin splitting torque~\cite{spin_splitter_torque1}, a crystal Hall effect~\cite{AHE1,AHE2,AHE3}, unique magneto-optical responses~\cite{MOE1,MOE2,MOE3}, and unconventional thermal transport~\cite{thermal}. Moreover, ATMs host chirally split magnons~\cite{ATM_magnon1,ATM_magnon2,ATM_magnon3,ATM_magnon4,ATM_magnon5,ATM_magnon6}---the quanta of spin waves~\cite{Magnon_1,Magnon_2,Magnon_3,Magnon_4,Magnon_5}---which are promising for low-dissipation spin transport~\cite{ATM_magnon2,ATM_magnon7,ATM_magnon8,ATM_magnon9}. Efficient magnon transport in conventional AFMs is often hindered by magnetization compensation, requiring a large magnetic field or strong spin-orbit coupling to overcome~\cite{Magnon_nonlocal_transport_1,Seebeck_effect,Magnon_nonlocal_transport_2,Magnon_transport_3,Nernst_effect,Magnon_nonlocal_transport_4,Magnon_nonlocal_transport_5,Magnon_nonlocal_transport_6,Magnon_nonlocal_transport_7,Magnon_nonlocal_transport_8,Magnon_transport_9,Magnon_transport_10,canted,Hanle_1,Hanle_2,Hanle_3,Electrical_injection1,Electrical_injection2}. ATMs intrinsically bypass this via chiral magnon splitting, enabling effects like the temperature-gradient-driven magnon spin Nernst effect~\cite{ATM_magnon2,ATM_magnon7,ATM_magnon9}, previously associated with magnon spin swapping in AFMs~\cite{Magnon_nonlocal_transport_8}.

Electrical spin injection from a heavy metal is a highly efficient method to generate magnon spin currents, demonstrated for longitudinal spin flow in ferromagnets~\cite{magnon_current_FM}, ferrimagnets~\cite{Electrical_injection0,detection_nonlocal}, and antiferromagnets~\cite{Hanle_1,Hanle_2,Hanle_3,Electrical_injection1,Electrical_injection2}. Its behavior in altermagnetism, however, remains wanting.

\begin{figure}[htbp]
    \centering
    \includegraphics[width=9.05cm, clip, trim=0.0cm 0cm 0cm 0cm]{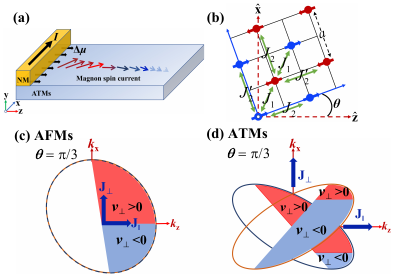}
    \caption{Diffusive magnon spin transport in ATMs $vs$ AFMs under electrical injection. (a) A charge current $I\hat{\bf x}$ in the normal metal (NM) creates a spin accumulation $\Delta{\pmb \mu}\parallel\hat{\bf z}$ via spin Hall effect, driving a vector magnon flow in ATMs. Red and blue arrows indicate the reversed transverse flow.  (b) Minimal ATMs model with the N\'eel vector ${\bf n}$ forming an angle $\theta$ to the diffusion $\hat{\bf z}$-axis. (c,d) Transverse magnon flow in AFMs and ATMs when ${\bf n}\nparallel \hat{\bf z}$. Red (blue) denotes positive (negative) transverse group velocities ${\bf v}_{\perp}\parallel \hat{\bf x}$ of injected magnons. }
    \label{fig1}
\end{figure}

In this Letter, we fill this gap by exploring the electrically injected, diffusion-driven transport of magnon spins in ATMs, developing a microscopic quantum kinetic equation, and performing a symmetry analysis. We demonstrate that the electron spin accumulation at a heavy-metal--ATM interface produces a ``vector" magnon spin current, i.e., multidirectional propagation containing both transverse and longitudinal spin flows [Fig.~\ref{fig1}(a) and (b)]. Notably, the transverse part reverses its direction when flowing away from the source due to \textit{intrinsic} different magnon diffusion lengths, and can be turned on or off by reorienting the N\'eel vector. Surprisingly, the transverse current also exists in conventional AFMs when the N\'eel vector tilts away from the diffusion axis, as dispersion anisotropy (especially at high frequency) distorts the group velocity distribution [Fig.~\ref{fig1}(c)]. However, in AFMs, the transverse velocities ${\bf v}_{\perp}$ of two magnon modes nearly compensate, yielding only a weak net transverse spin current. In ATMs, by contrast, magnon spin splitting from broken parity-time symmetry creates strong asymmetry between the two modes, leading to a large imbalance in transverse velocity and a greatly enhanced transverse spin current [Fig.~\ref{fig1}(d)]. This intrinsic mechanism enhances the net transverse response in ATMs by two orders of magnitude compared to AFMs of similar crystal symmetry, offering a clear transport signature to distinguish ATMs from conventional AFMs.

We model the ATMs using a minimal square-lattice model with lattice constant $a$, shown in Fig.~\ref{fig1}(b). The spin Hamiltonian reads
\begin{align}
    \hat{H}&=J_1\sum_{\langle i,j\rangle}\hat{\bf S}_{1,i}\cdot\hat{\bf S}_{2,j}-\gamma\hbar\sum_{i}{\bf H}_0\cdot(\hat{\bf S}_{1,i}+\hat{\bf S}_{2,i})\nonumber\\
    &+J_2\sum_{\langle i_z,j_z\rangle}\hat{\bf S}_{1,i}\cdot\hat{\bf S}_{1,j}+J_2'\sum_{\langle i_x,j_x\rangle}\hat{\bf S}_{1,i}\cdot\hat{\bf S}_{1,j}\nonumber\\
    &+J_2'\sum_{\langle i_z,j_z\rangle}\hat{\bf S}_{2,i}\cdot\hat{\bf S}_{2,j}+J_2\sum_{\langle i_x,j_x\rangle}\hat{\bf S}_{2,i}\cdot\hat{\bf S}_{2,j}\nonumber\\
    &+K\sum_{i}(\hat{\bf S}_{1,i}^z)^2+K\sum_{i}(\hat{\bf S}_{2,i}^z)^2.
    \label{Hamiltonian_atm}
\end{align}
Here, $\hat{\bf S}_{1,i}$ and $\hat{\bf S}_{2,j}$ represent spins of magnitude $S$ on \textit{occupied} lattice sites of sublattices 1 and 2. $\langle i,j\rangle$ denotes the nearest-neighboring coupling between different sublattices along the diagonal direction of the square lattice, while $\langle i_x,j_x\rangle$ and $\langle i_z,j_z\rangle$ indicate the nearest-neighboring coupling of the same sublattice along two main axis of the crystal. $J_1>0$ is the antiferromagnetic inter-sublattice exchange constant, $J_2~(J_2')$ is the intra-sublattice exchange constant, $K<0$ is uniaxial anisotropy constant, $\gamma$ is the electron gyromagnetic ratio, and ${\bf H}_0$ is the applied magnetic field aligned with the N\'eel vector ${\bf n}$.  The N\'eel vector is tunable and forms an angle $\theta$ with the $\hat{\bf z}$-direction [Fig.~\ref{fig1}(b)].

The dispersions of modes 1 and 2  are chirally split, carrying opposite spins $-\hbar$ and $+\hbar$ along the N\'eel vector (see Supplemental Material (SM)~\cite{supplement} for details):
\begin{align}
    \hbar\omega_{1/2,{{\mathbf k}}}(\theta)&=\pm S\left(J_2-J_2'\right)\left(\gamma_1({{\mathbf k}})-\gamma_2({{\mathbf k}})\right)\nonumber\\&+{4J_1S\gamma_3({{\mathbf k}})\Delta_{{\mathbf k}}}/{\gamma_e({{\mathbf k}})}\pm\gamma\hbar H_0,
    \label{dispersion}
\end{align}
in which $\gamma_1({\bf k})=\cos\left[\left(k_x\sin{\theta}+k_z\cos{\theta}\right)a\right]$, $\gamma_{2}({\bf k})=\cos [( \left(k_x\cos{\theta}-k_z\sin{\theta}\right)a]$, and $\gamma_3({\bf k})=\cos[(k_x\sin{\theta}+k_z\cos{\theta})a/2] \cos\left[\left(k_x\cos{\theta}-k_z\sin{\theta}\right)a/2\right]$ are the form factors, $\gamma_e({\mathbf k})=4J_1\gamma_3({\mathbf k})/[4J_1-2K+(J_2+J_2')(\gamma_1({\mathbf k})+\gamma_2({\mathbf k})-2)]$,  and $\Delta_{\mathbf k}=\sqrt{1-\gamma_e^2({\mathbf k})}$. The magnon chirality splitting vanishes when $J_2=J_2'$, reducing to the conventional AFMs.

We develop the microscopic quantum kinetic equations to account for the electrical-injection, coherence, diffusion, and dephasing of magnon spin current in ATMs$/$AFMs in the geometry Fig.~\ref{fig1}(a).
Magnons in the ATMs couple with the conduction electrons in normal metal (NM) via the interfacial $s$-$d$ exchange interaction~\cite{s-d_interaction_1,s-d_interaction_2,s-d_interaction_3,s-d_interaction_4,canted} $\hat{H}_{\rm int}={\cal J}a^3\sum_{l}\sum_{i\in{\rm int}}^{N}\int d{\bf r}\delta({\bf r}-{\bf R}_{i})\hat{\bf s}({\bf r})\cdot\hat{\bf S}_{l,i}$,
where ${\cal J}$ is the exchange coupling strength, $\hat{\bf s}({\bf r})$ is the electron spin density, and $N$ is the number of interface sites. This interaction leads to the interfacial scattering between electron operator $\{\hat{c}^{\dagger}_{{\bf k}\zeta},\hat{c}_{{\bf k}\chi}\}$ and magnon operator $\hat{b}_{\xi,{\bf q}}$ according to  
\begin{align}
        \hat{H}_{\rm int}&=\frac{a}{d}\sqrt{\frac{S}{2N}}{\cal J}\sum_{\zeta,\chi=\uparrow,\downarrow}\sum_{{\bf k},{\bf k}'}\sum_{\bf q}\sum_{\xi=1,2}\left({\pmb \sigma}_{\zeta\chi}\cdot \delta{\bf S}_{{\bf q}}^{\xi}\right)\nonumber\\
        &\times\hat{c}_{{\bf k}'\zeta}^{\dagger}\hat{c}_{{\bf k}\chi}\hat{b}_{\xi,{\bf q}}\delta_{{\bf k}+{\bf q},{\bf k}'}+\text{H.c.},
        \label{H_int}
\end{align}
where $d$ is the thickness of NM and ${\pmb \sigma}$ is the Pauli matrices. The scattering amplitude for electron–magnon processes is captured by the kernel
\begin{align}
    {\pmb{\sigma}}\cdot\delta{\bf S}_{\bf q}^{\xi=1,2}&\sim\left(\begin{array}{cc}
        -\sin{\theta} & \cos{\theta}\mp1 \\
        \cos{\theta}\pm1 & \sin{\theta}
    \end{array}\right),
    \label{SigmaDeltaS}
\end{align}
which depends on the orientation $\theta$ of the N\'eel vector ${\bf n}$.
When $\theta=0$ (i.e., ${\bf n}\parallel \hat{\bf z}$), only the off-diagonal $\zeta\neq\chi$ elements are nonzero in Eq.~\eqref{SigmaDeltaS}, indicating that magnon scattering induces spin flips of the electrons—for instance, absorption of a mode‑1 magnon flips an electron from spin‑up to spin‑down. When $\theta=\pi/2$ (${\bf n}\perp \hat{\bf z}$), all the matrix elements in Eq.~\eqref{SigmaDeltaS} present with the same magnitude, allowing both spin‑conserving and spin‑flip processes to occur and interfere. This interference ultimately suppresses the magnon spin current (see SM~\cite{supplement} and the calculation below). For an arbitrary $\theta$, this interference is not balanced, leading to a vector spin current with multidirectional flow.

We verify these expectations by modeling the magnon injection driven by a spin accumulation $\Delta\mu=\delta\mu_{\uparrow}-\delta\mu_{\downarrow}$ polarized along $\hat{\bf z}$.
Under electrical injection and dissipation, the magnon density matrix ${\rho}_{\bf q}$, in which the diagonal terms represent magnon population and the off-diagonal terms denote the mode correlation~\cite{canted}, obeys the quantum kinetic equation
$\partial_t\rho_{\bf q}=\partial_t\rho_{\bf q}|_{\text{inj}}+\partial_t\rho_{\bf q}|_{\text{coh}}+\partial_t\rho_{\bf q}|_{\text{rex}}$. The magnon dispersion ${\cal E}_{\bf q}=\text{diag}\left(\hbar\omega_{1,{\bf q}},\hbar\omega_{2,{\bf q}},\hbar\omega_{1,-{\bf q}},\hbar\omega_{2,-{\bf q}}\right)$ drives the coherent dynamics of non-equilibrium magnons according to $\partial_{t}\rho_{\bf q}|_{\text{coh}}=(-i/\hbar)(\rho_{\bf q}\sigma_3{\cal E}_{\bf q}-{\cal E}_{\bf q}\sigma_3\rho_{\bf q})$, where $\sigma_3=\text{diag}(1,1,-1,-1)$ is the metric. Within the relaxation-time approximation~\cite{relaxation-time1,relaxation-time2}, the magnons relax toward to their equilibrium $\rho_{\bf q}^{(0)}$ by
$\partial_t\rho_{\bf q}|_{\rm rex}=-(\rho_{\bf q}-\rho_{\bf q}^{(0)})/\tau_m$ with a characteristic time $\tau_m$
assumed equal for both modes~\cite{costa2025giant}.
The spin accumulation drives the electrical injection as described by
\begin{align}
    &\partial_{t}\rho_{\bf q}|_{\rm{inj}}=\frac{\pi}{\hbar}\frac{S}{2N}\frac{a^2}{d^2}{\cal J}^2\sum_{\bf k}\sum_{\zeta,\chi=\uparrow,\downarrow}\Big[f_{{\bf k+q},\zeta}\left(1-f_{{\bf k},\chi}\right)(\sigma_3+\rho_{\bf q})\nonumber\\
    &-f_{{\bf k},\chi}\left(1-f_{{\bf k+q},\zeta}\right){\rho}_{\bf q}\Big]{\cal M}_{\zeta\chi}({\bf k},{\bf k+q})+\text{H.c.},
    \label{injection_equation}
\end{align}
where $f_{\bf k,\zeta}=\left[f_{\bf k}^0-(\partial{f_{\bf k}^{0}}/\partial\varepsilon_{\bf k})\delta\mu_{\zeta}\right]$ is its non-equilibrium  distribution for spin-$\zeta$ (shifted from the equilibrium Fermi-Dirac distribution $f_{\bf k}^0$), $\varepsilon_{\bf k}$ is the electron kinetic energy, 
and ${\cal M}_{\zeta\chi}({\bf k},{\bf k}+{\bf q})$ is a $4\times4$ matrix encoding all the electron-magnon scattering processes in Eq.~\eqref{SigmaDeltaS} (see SM~\cite{supplement} for details).

The injected magnons around the source located at $z=0$ in the setup Fig.~\ref{fig1}(a) diffuse along $\hat{\bf z}$. Only the diagonal components $\rho_{11}({\bf q})$ and $\rho_{22}({\bf q})$ of ${\rho_{\bf q}}$ carry spins, which diffuse when away from the source according to
\begin{align}
    -{v}_{\eta}^{\parallel}({\bf q})\frac{\partial\rho_{\eta\eta}(z,{\bf q})}{\partial{z}}-\frac{\rho_{\eta\eta}(z,{\bf q})-\rho_{\eta\eta}^{(0)}({\bf q})}{\tau_m}=0,
    \label{diffusion_equation}
\end{align}
where $\eta=\{1,2\}$ indicate the modes. Solving Eq.~\eqref{diffusion_equation} yields the magnon distribution at the region $z>0$: when $\quad v_{\eta}^{\parallel}({\bf q})>0$, $\rho_{\eta\eta}(z,{\bf q})= \rho_{\eta\eta}^{(0)}({\bf q}) 
+ [\rho_{\eta\eta}({\bf q})-\rho_{\eta\eta}^{(0)}({\bf q})]e^{-z/(v_{\eta}^{\parallel}({\bf q})\tau_m)}$; when $v_{\eta}^{\parallel}({\bf q})<0$, $\rho_{\eta\eta}(z,{\bf q}) = \rho_{\eta\eta}^{(0)}({\bf q})$.
They contribute to the spin-current density
\begin{align}
    {J}_{i}^{\alpha}(z)&=\sum_{\eta=1,2}\sum_{\bf q}\Big({\pmb {\cal S}}^{\alpha}_{\eta}({\bf q})\otimes{\bf v}^{i}_{\eta}({\bf q})\rho_{\eta\eta}(z,{\bf q})\Big),
    \label{current}
\end{align}
where the superscript $\alpha=\{x,z\}$ is the spin-polarization direction, the subscript $i=\{\parallel,\perp\}$ represents the longitudinal/transverse flow direction, and ${\pmb {\cal S}}_{1/2}({\bf q})=\mp\sin{\theta}\hat{\bf x}\mp\cos{\theta}\hat{\bf z}$ is the magnon spin.

The calculation is performed using realistic material parameters: $J_1=1~\text{meV}$, $J_2=-0.15~\text{meV}$, $J_2'=0.5~\text{meV}$, $K=-0.02~\text{meV}$, $S=3/2$, lattice constant $a=1~\text{nm}$, possible for Cr$_2$I$_2$O and Cr$_2$Br$_2$O~\cite{ATM_magnon2}, and NM thickness $d=20~\text{nm}$~\cite{Magnon_nonlocal_transport_4,Electrical_injection1}. Similar phenomena may be expected in other candidates, e.g., LuFeO$_3$~\cite{ATM_magnon9} and MnTe~\cite{spin_splitting_2,MnTe1}. 
The other parameters are chosen as $\Delta\mu=0.4$~meV~\cite{s-d_interaction_1,spin_accumulation1,spin_accumulation2,spin_accumulation3}, relaxation time $\tau_m=1~\text{ns}$, and interfacial exchange coupling ${\cal J}=15~\text{meV}$~\cite{s-d_interaction_2,s-d_interaction_3,Strong_inter1,Strong_inter2}.

\begin{figure}[htbp]
    \centering
    \includegraphics[width=8.7cm]{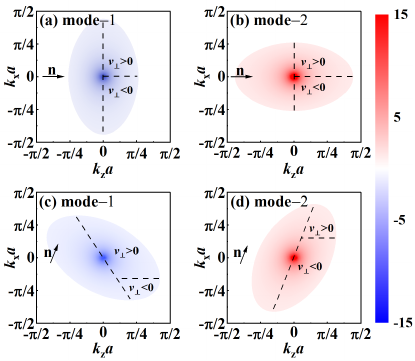}
    \caption{Injected population of modes-1 and 2 magnons under different N\'eel-vector directions $\theta=\{0,\pi/3\}$.}
    \label{fig2}
\end{figure}

Recall that a temperature gradient drives no magnon spin current in AFMs without applied magnetic fields due to the magnetization compensation~\cite{Magnon_nonlocal_transport_1,Nernst_effect,Magnon_nonlocal_transport_5,Magnon_nonlocal_transport_7}. Remarkably, however, we find electric injection generates a longitudinal spin current in AFMs when ${\bf n}\parallel\hat{\bf z}$ free of magnetic fields (taking $J_2=J_2'$). This is because the electric drive absorbs mode-1 magnons while injecting mode-2 magnons. Figure~\ref{fig2}(a) and (b) illustrate this magnon distribution in the Brillouin zone for ATMs, in which different group velocities of two modes further enhance the spin current. When the N\'eel vector deviates from the diffusion $\hat{\bf z}$-direction, e.g., at $\theta = \pi/3$, the injected magnon distribution becomes tilted, as shown in Fig.~\ref{fig2}(c) and (d). Only injected magnons with ${\bf v}_{\parallel}>0$ contribute to the spin current at $z>0$. Among these, the proportions with ${\bf v}_{\perp}>0$ and ${\bf v}_{\perp}<0$ become unbalanced: more magnons with ${\bf v}_{\perp}>0$ are absorbed for mode-1 [Fig.~\ref{fig2}(c)], while more with ${\bf v}_{\perp}<0$ are injected for mode-2 [Fig.~\ref{fig2}(d)]. This imbalance produces a large transverse magnon spin flow. When $\theta=\pi/2$, Eq.~\eqref{SigmaDeltaS} implies that all electron–magnon scattering amplitudes are equal in magnitude, so spin-conserving and spin-flip processes occur with equal probability and interfere, resulting in zero net magnon injection for both modes (see SM~\cite{supplement}).

\begin{figure}[htbp]
    \centering
    \includegraphics[width=8.7cm]{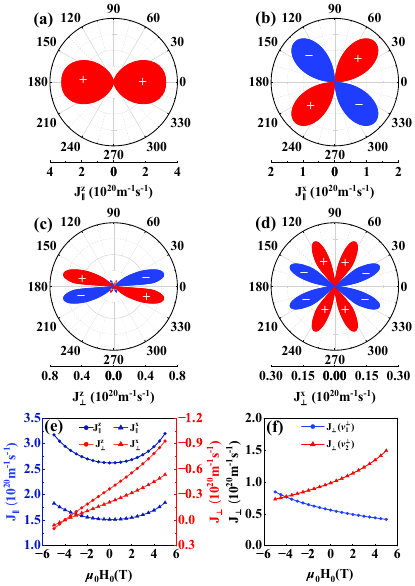}
    \caption{Dependence of longitudinal $J_{\parallel}^{z,x}$ [(a,b)] and transverse $J_{\perp}^{z,x}$ [(c,d)] magnon spin currents on the N\'eel-vector direction $\theta$ at $z\gtrsim 0$. (e) and (f) illustrate efficient control of these spin currents by the applied magnetic field at $\theta = \pi/6$.}
    \label{fig3}
\end{figure}

This dependence of electrical injection on the N\'eel vector orientation governs the magnon spin flow below the NM at $z\gtrsim 0$, as summarized in Figs.~\ref{fig3}(a)-(d). The longitudinal components $J_{\parallel}^{z}$ and $J_{\parallel}^{x}$ exhibit two-fold and four-fold symmetries, while the transverse components $J_{\perp}^{z}$ and $J_{\perp}^{x}$ display double four-fold symmetries.  All components vanish when ${\bf n}\perp\hat{\bf z}$ due to vanishing magnon injection, and the transverse components vanish for ${\bf n}\parallel\hat{\bf z}$ because transverse velocities cancel [Fig.~\ref{fig2}(a) and (b)].  Transverse components also vanish at $\theta=\{\pi/4,3\pi/4,5\pi/4,7\pi/4\}$, where contributions from the two magnon modes precisely offset each other. Crucially, the AFMs exhibit similar features but with a strongly suppressed transverse current. Therefore, clear experimental evidence rests on the transverse-to-longitudinal ratio, which is two orders of magnitude larger in ATMs than in AFMs. Electrical injection exhibits symmetries distinct from longitudinal ($A_3\cos^2\theta+B_3\sin^2\theta$) and transverse ($\sin\theta\cos\theta$) spin currents driven by a temperature gradient.

This clear dependence on the N\'eel-vector orientation of the vector magnon spin flow suggests that its generation and transport under electrical injection beneath the heavy metal is governed by symmetry and dynamical constraints. The magnon spin flow $J_{i}^{\alpha}$ near the heavy metal is related to the spin-accumulation gradient $\delta\mu_{j}^{\beta}$ of electrons with gradient $j$-direction and spin-polarization $\beta$-direction via the fourth-order tensor $\Gamma_{i j}^{\alpha \beta}$ according to 
\begin{align}
J_{i}^{\alpha}=\Gamma_{i j}^{\alpha \beta}\delta\mu_{j}^{\beta}.
\end{align}
In $J_{i}^{\alpha}$ and $\Gamma_{i j}^{\alpha \beta}$, the subscripts $\{i, j\} \in \{x, z\}$ denote flow directions, while the superscripts $\{\alpha, \beta\} \in \{x, z\}$ specify spin-polarization orientations. In the geometry Fig.~\ref{fig1}(a), $\beta=z$ is the spin-polarization direction, while $j=z$ is the gradient direction of the spin accumulation. Our microscopic calculation reveals that all components of $\Gamma_{ij}^{\alpha\beta}$ are allowed in both ATMs and AFMs, which implies that a transverse magnon spin current flowing along $x$ can emerge in either system.

As the N\'eel vector ${\bf n}$ rotates in the film $xz$-plane, the magnon spin current evolves accordingly. Consequently, the response tensor $\Gamma_{ij}^{\alpha\beta}$ is assumed to be expanded as a series in the components of ${\bf n}=(n_x,0,n_z)$~\cite{EPM1,EPM2}. Because $J_{i}^{\alpha}$ is invariant under the N\'eel-vector reversal (${\bf n}\rightarrow-{\bf n}$), only even-order terms appear in the expansion. A further dynamical constraint requires the magnon spin current to vanish when ${\bf n}$ is perpendicular to the spin-accumulation polarization. For interpreting, we consider two mirror operations ${\cal M}_x$ and ${\cal M}_z$ with respect to the $x$-$y$ and $y$-$z$ planes. The transformation of the response tensor and the N\'eel vector is summarized in Table~\ref{transformations} (see SM~\cite{supplement} for details).
Subsequent symmetry operations determine the explicit dependence of the tensor components on ${\bf n}$, as summarized in Table~\ref{symmetry_chars} (see SM~\cite{supplement} for details).

\begin{table}[htp!]
\centering
\caption{Transformation of  response tensor $\Gamma_{ij}^{\alpha\beta}$ and N\'eel vector under mirror operations $\{{\cal M}_x,{\cal M}_z\}$.}
\label{transformations}
\setlength{\tabcolsep}{27pt}
\begin{tabular}{c|c|c}
\hline
elements & ${\cal M}_x$ & ${\cal M}_z$ \\
\hline
$\Gamma_{zz}^{zz}$ & $\Gamma_{zz}^{zz}$ & $\Gamma_{zz}^{zz}$ \\
$\Gamma_{zz}^{xz}$ & $-\Gamma_{zz}^{xz}$ & $-\Gamma_{zz}^{xz}$ \\
$\Gamma_{xz}^{zz}$ & $-\Gamma_{xz}^{zz}$ & $-\Gamma_{xz}^{zz}$ \\
$\Gamma_{xz}^{xz}$ & $\Gamma_{xz}^{xz}$ & $\Gamma_{xz}^{xz}$ \\
$n_x$ & $-n_x$ & $n_x$ \\
$n_z$ & $n_z$ & $-n_z$ \\
\hline
\end{tabular}
\end{table}

Microscopic model calculations yield the expansion coefficients: while the longitudinal coefficients are nearly identical between ATMs and AFMs ($A_1\approx A_2$, $B_1\approx B_2$), the transverse coefficients in ATMs are strongly enhanced, exceeding those in AFMs by almost two orders of magnitude ($C_{1}\approx40C_2$, $D_{1}\approx40D_2$, and $E_1\approx30E_2$).

\begin{table}[htp!]
\centering
\caption{Dependence of response tensor $\Gamma_{ij}^{\alpha\beta}$ on the N\'eel vector $(n_x,0,n_z)=(\sin{\theta},0,\cos{\theta})$ in ATMs and AFMs. $\{A_{1,2}, B_{1,2}, C_{1,2}, D_{1,2}, E_{1,2}\}$ are expansion coefficients for ATMs and AFMs.}
\label{symmetry_chars}
\setlength{\tabcolsep}{8pt}
\begin{tabular}{ccc}
\toprule
 & ATMs & AFMs \\
\midrule
$\Gamma_{zz}^{zz}$ & $A_1\cos^2\theta$ & $A_2\cos^2\theta$ \\
$\Gamma_{zz}^{xz}$ & $B_1\sin\theta\cos\theta$ & $B_2\sin\theta\cos\theta$ \\
$\Gamma_{xz}^{zz}$ & $C_1\sin 2\theta + D_1\sin 4\theta$ & $C_2\sin 2\theta + D_2\sin 4\theta$ \\
$\Gamma_{xz}^{xz}$ & $E_1\,(\cos 2\theta - \cos 6\theta)$ & $E_2\,(\cos 2\theta - \cos 6\theta)$ \\
\bottomrule
\end{tabular}
\end{table}

The magnetic field along the N\'eel vector efficiently controls spin currents [Fig.~\ref{fig3}(e)].
The longitudinal components $J_{\parallel}^{z,x}(H_0)\approx J_{\parallel}^{z,x}(-H_0)$ are nearly symmetric about $H_0=0$, while the transverse components are asymmetric. This asymmetry arises from opposite frequency shifts of two magnon modes: increasing $H_0$ up-shifts mode-1 frequency, reducing magnon absorption, while down-shifting mode-2 frequency enhances magnon generation, and vice versa. Therefore, the magnon current carried by the two modes, $J_{\perp}(v^{\perp}_1)$ and $J_{\perp}(v^{\perp}_2)$, depends oppositely on the magnetic field [Fig.~\ref{fig3}(f)], thus explaining the $H_0$-dependence of net transverse current $J_{\perp} = J_{\perp}(v^{\perp}_1) - J_{\perp}(v^{\perp}_2)$.

\begin{figure}[htp]
    \centering
    \includegraphics[width=8.7cm]{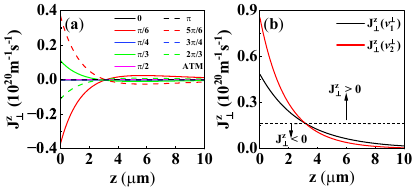}
    \caption{(a) Transverse spin current $J_{\perp}^{z}$ as a function of the propagation distance $z$ for different orientations $\theta$ of the N\'eel-vector. (b) Spatial profiles of the mode-resolved contributions $J_{\perp}^{z}(v^{\perp}_1)$ and $J_{\perp}^{z}(v^{\perp}_1)$ at $\theta = \pi/6$.}
    \label{fig4}
\end{figure}

The spatial evolution of the transverse spin current $J_{\perp}^{z}$ exhibits a nontrivial sign reversal at a finite distance, as shown in Fig.~\ref{fig4}(a), indicating a reorientation of the vector current. This behavior originates from the unequal decay rates of the two magnon modes. For example, at $\theta=\pi/6$ [Fig.~\ref{fig4}(b)], the contribution of mode-1 $J_{\perp}^{z}(v^{\perp}_1)$ is smaller at $z=0$ but decays slowly, while $J_{\perp}^{z}(v^{\perp}_2)$ is larger initially but decays rapidly. Consequently, the net transverse current reverses at a finite propagation distance.

In summary, we introduce the concept of ``Vector Magnonics" by showing that electrical spin injection drives a vector magnon spin current---with both longitudinal and transverse components---in ATMs and AFMs, which can laterally connect different spintronic units on a chip. This current is electrically controllable via the N\'eel vector or spin-accumulation orientation.  A key quantum signature is the spatial reversal of the transverse flow direction, arising from interference and unequal decay between two chirally split magnon modes. Crucially, our combined microscopic and symmetry analysis reveals that the transverse-to-longitudinal spin-current ratio is enhanced by two orders of magnitude in ATMs compared with AFMs due to broken parity-time symmetry.  Thus, this giant ratio provides a decisive experimental fingerprint to distinguish ATMs from AFMs, offering a much-needed transport probe for this emerging field.

\begin{acknowledgments}
This work is financially supported by the National Key Research and Development Program of China under Grant No.~2023YFA1406600 and the National Natural Science Foundation of China under Grants No.~12374109 and 12474100. 
\end{acknowledgments}

\end{document}